# Atomic scale analysis of grain boundary deuteride growth front in Zircaloy-4

A.J. Breen[a,*], I. Mouton[a], W. Lu[a], S. Wang[b], A. Szczepaniak[a], P. Kontis[a], L.T. Stephenson[a], Y. Chang[a], A.K. da Silva[a], C. Liebscher[a], D. Raabe[a], T.B. Britton[b], M. Herbig[a], B. Gault[a]

[a] Max-Planck-Institut für Eisenforschung, Max-Planck-Straße 1, 40237 Düsseldorf, Germany.
[b] Department of Materials, Royal School of Mines, Imperial College London, London, SW7 2AZ, UK

*corresponding author
*E-mail address*: a.breen@mpie.de (A.J. Breen)
*Postal address:* Max-Planck-Institut für Eisenforschung, Max-Planck-Straße 1, 40237 Düsseldorf, Germany.

**Abstract:**

Zircaloy-4 (Zr-1.5%Sn-0.2%Fe-0.1%Cr wt. %) was electrochemically charged with deuterium to create deuterides and subsequently analysed with atom probe tomography and scanning transmission electron microscopy to understand zirconium hydride formation and embrittlement. At the interface between the hexagonal close packed (HCP) $\alpha$-Zr matrix and a face centred cubic (FCC) $\delta$ deuteride ($ZrD_{1.5-1.65}$), a HCP $\zeta$ phase deuteride ($ZrD_{0.25-0.5}$) has been observed. Furthermore, Sn is rejected from the deuterides and segregates to the deuteride/$\alpha$-Zr reaction front.

Zircaloy-4 is primarily used as a fuel cladding material in water based nuclear reactors due to its low neutron absorption cross section, good mechanical properties and corrosion resistance. However, the alloy is susceptible to hydrogen embrittlement (HE) through a mechanism known as delayed hydride cracking (DHC) or hydrogen-induced delayed cracking (HIDC) [1, 2]. Hydrogen ingress occurs during service due to exposure with the water coolant. Zirconium has a relatively high oxidation potential and will readily form a $ZrO_2$ layer when in contact with water – through this process, hydrogen is released, some of which is absorbed into the underlying alloy [3]. The terminal solid solubility (TSS) of hydrogen in $\alpha$-Zr is low, less than 10 wt. ppm at room temperature and ambient pressure [4], and so there is a high driving force for excess hydrogen to precipitate as a brittle solid zirconium hydride. The severity of the embrittlement is dependent on several factors, for instance, an increase in the volume fraction and size of the hydrides and a decrease in the strain rate are known to be more deleterious [2, 5, 6]. The orientation of hydride platelets also has a significant influence on toughness, with those perpendicular to the tensile-axis being more deleterious [7]. The atomic structure and interfacial chemistry are also expected to play a critical role. In particular it is interesting to note that Northwood et al. [8] report that during DHC, crack initiation occurs within the hydride itself and not at the hydride/α interface. It therefore follows that knowledge of the internal structure and chemistry is of utmost importance to elucidating the mechanisms of DHC.

However, despite many decades of research on hydrogen interaction in zirconium alloys, including Zircaloy-4, there is still a lack of atomic-scale structural and compositional characterisation of the hydrides including the growth front, which is a focus in this letter. Significant uncertainty also exists in the thermodynamics of the Zr-H system and variation in the proposed phase diagrams is evident, particularly below the eutectic temperature of 550 ˚C [3, 9, 10]. Currently, four different zirconium hydride phases have been reported. In terms of crystal structure and approximate stoichiometry $ZrH_X$, these are referred to as hexagonal close packed (HCP) $\zeta$ (x=0.25-0.5), face-centred tetragonal (FCT) $\gamma$ (x=1), face-centred cubic (FCC) $\delta$ (x≈1.5-1.65) and FCT $\varepsilon$ (x≈1.75-2) [11-13]. $\zeta$ is fully coherent with the $\alpha-Zr$ matrix with a c-axis twice as long, and may play a critical role as the intermediate phase in the dissolution and precipitation processes of the more stable $\delta$-hydride [14, 15]. The ambiguity in identifying the various hydride phases, particularly $\zeta$, has been brought about, in part, by the much lower diffraction intensities of the hydride relative to the $\alpha-Zr$ matrix [14]. However, with the advent of aberration corrected transmission electron microscopy (TEM), greater clarity of atomic structure is now possible [16].

Atom probe tomography (APT) is another powerful microanalytical technique that offers a unique opportunity to reconstruct the three-dimensional position and elemental identity of atoms from a material specimen with sub-nm spatial resolution [17]. The technique has already been used to observe solute segregation, including H, at grain boundaries and the metal/oxide interface in zirconium alloys [18-22] but has had limited application to the study of zirconium hydrides directly. This is likely partially due to the technical challenges involved in identifying and quantifying hydrogen using APT. Background H from the analysis chamber, typically in the range of 0.5-5 at.% [23], is routinely observed in the mass spectra. Ambiguity arises in determining which hydrogen atoms are from the chamber or the sample being analysed. Yet, recently, Chang et al. [24], reported on successfully using APT to characterise hydrides in the thermodynamically similar system of Ti-H – here we demonstrate that it can also be used to gain insights into the Zr-H system as well. A way to mitigate the challenges associated to H-quantification, as employed in this study, is to charge the material with the isotope deuterium (D or $^2H$), to minimise overlap with the background hydrogen peaks [23, 25-27]. As a consequence, H analysis in atom probe is starting to emerge as a promising means of directly observing hydrogen in metals. Coupled with aberration corrected TEM as well as electron channelling contrast imaging (ECCI) [28] and electron back-scattered diffraction (EBSD) in the scanning electron microscope (SEM) – a multiscale approach to zirconium hydride analysis is presented herein.

Commercial Zircaloy-4 (Zr-1.5%Sn-0.2%Fe-0.1%Cr wt. %) was received as a rolled and recrystallized plate with a typical split basal texture and average grain size of ~ 11 μm. The sample was then heat treated at 800 ˚C for two weeks to form large 'blocky-alpha' grains > 200 μm similar to that recently reported by Tong and Britton [29]. The sample was then electrochemically charged with deuterium (galvanostatic charging, current density = 2 kA/m$^2$) using a solution of 1.5 wt. % $D_2SO_4$ in $D_2O$ at 65 ˚C for 24 hours. After this process, a hydride layer of approximately 20 μm thickness formed at the surface. Annealing at 400 ˚C for 5 hours followed by furnace cooling of 0.5 ˚C/min was then used to redistribute the D from the surface into the bulk of the sample. The microstructure was then inspected using polarised light microscopy and a Zeiss Merlin scanning electron microscope (SEM).

Fig. 1 provides an overview of the microstructure after D charging and annealing. Fig. 1 a is a polarised light-optical micrograph showing the different types of deuterides that have evolved. The grain boundary deuterides are the most common and the focus of further characterisation. Fig. 1 b is an ECCI image of a deuteride at higher magnification. Here, contrast changes are observed within the hydride and suggests a high dislocation density and internal strain as well as possible phase separation. The complex nature of the contrast in ECCI does not allow for directly assessing the size and nature of the various phases, but what seems like a single deuteride appears to be complex in nature and will be hereafter referred to as a deuteride packet. Fig. 1 c is an EBSD IPF map of the same region, the deuteride is indexed as δ and follows the $\{0001\}_\alpha \parallel \{111\}_\delta$ , $<11\text{-}20>_\alpha \parallel <110>_\delta$ orientation relationship with the bottom matrix grain. This is in agreement with previous reports [14, 30]. Fig. 1 d is a misorientation to grain average orientation map which suggests an orientation gradient within the deuteride packet that is associated with a high dislocation density and internal strain. This deuteride packet also likely contains other hydrides, e.g. ζ, as well as metallic inclusions, but with respective volume fractions that do not allow for direct imaging at this scale, hence TEM was pursued.

An overview of the TEM results is shown in Fig. 2. A TEM specimen was hence prepared from within the same grain boundary deuteride packet using a FEI Helios 600i dual-beam scanning electron microscope / focused-ion beam (SEM/FIB). The FIB was used to lift out a section of the grain boundary deuteride, as indicated in Fig. 2 a, and was subsequently thinned until it was electron transparent. Scanning transmission electron microscopy (STEM) imaging was conducted in an aberration-corrected STEM/TEM (FEI Titan Themis) at 300 kV. For high-resolution high-angle annular dark field image (HAADF) imaging, a probe semi-convergence angle of 17 mrad and inner and outer semi-collection angles ranging from 73 to 200 mrad were chosen. Fig. 2 b is a high-angle annular dark field image (HAADF) of the grain boundary deuteride growth front from this sample. Within the α-Zr grain, 100–200 nm ahead of the growth front, variations in contrast suggest that intragranular deuterides are formed. Within approx. 100 nm from the growth front, a deuteride depletion zone is readily visible. Variations in contrast are observed behind the growth front, in what is referred to as a transition zone between the α-Zr matrix into the δ-deuteride indicated in Fig. 2 b. This particular contrast could be partly due to an inclined interface relative to the incident electron beam, or to a rough interface. Fig. 2 c is a more highly magnified micrograph of this transitional zone that shows a rough interfacial network of 3 different phases. In Fig. 2 d the experimental fast Fourier transforms (FFT) computed from the highlighted regions in Fig. 2 c are contrasted with simulated electron diffraction patterns (using JEMS software [31]) of the suspected phases. The change in crystal structure across the growth front is clearly visible in the diffractograms. The slight deviations between the computed FFTs and the simulated diffraction patterns are related to dynamical scattering effects, local changes in sample thickness or orientation and variation in deuteride stoichiometry, but are considered negligible for phase identification. In particular, the additional sub-lattice spots, allow to unambiguously identify the ζ phase present in this transition zone.

The results from a representative APT dataset are summarised in Fig. 3. The APT specimens were prepared from a similar grain boundary deuteride to that shown in the previous figures, and a SEM micrograph of the deuteride is shown in Fig. 3 a. APT specimens were prepared using the lift-out procedure introduced by Thompson et al. [32] on an FEI Helios dual-beam plasma focused ion beam (PFIB). APT experiments were performed on a CAMECA local electrode atom probe (LEAP) 5000 XR. While voltage pulsing has typically been used in the past for H analysis related experiments in order to mitigate H surface diffusion and molecular ion evaporation [25-27], this was found to have very low success rates, while not entirely preventing the evaporation of molecular H species in this system. UV laser pulsing was thus used, which drastically improved yield. A base temperature of 60 K, laser pulse energy of 60 pJ, pulse repetition rate of 250 kHz and a detection rate of 1-2 ions per 100 pulses were used during the experiments.

A specimen containing the deuteride/alpha growth front, is shown in Fig. 3 b. In Fig. 3 c are displayed the atom maps of the reconstructed data for the Zr and Sn. From these maps, an interface between α-Zr and the selected grain boundary deuteride is readily visible. Sn is seen to partition to α-Zr with a segregate region in the vicinity of this interface. The presence of deuterium in the specimen can be clearly assessed by the peaks present in the mass spectrum displayed in Fig. 3 d, that shows a comparison between a loaded specimen and the one obtained from a control sample that had not been electrochemically loaded. In addition, a few peaks corresponding to H- and D- containing molecular ions, e.g. $ZrH^{2+}/ZrD^{2+}$ were also observed. The presence of hydrogen in the analysis may be due to a partial hydration during deuterium charging, the subsequent exchange of D by H or residual H from the APT chamber [24]. FIB milling may also induce hydride formation [33, 34] but the relative amount of D detected is consistent with the presence of a deuteride packet formed during the deuterium loading that has remained stable during sample preparation. Nevertheless, the distribution of D and precise quantification of the deuteride packet remains a challenge and an in-depth discussion on this topic falls outside the scope of this study.

Fig. 4 a shows a close-up on this interface taken from the region of interest (ROI) highlighted by the dashed line square in Fig. 3 c. This map highlights the variation in D-content (ions corresponding to the first peak in the mass spectrum, which is entirely H, have not been included in this analysis). Zr and Sn isocomposition surfaces have been used to highlight the rough ζ/α interface. A representative 1D composition profile across the interface is shown in Fig. 4 b. The variation in the deuterium content from δ into ζ and then α agrees qualitatively with our TEM observations. The increase in the amount of D at the δ/ζ and ζ/α interfaces could be due to changes in the local electrostatic field that leads to variations in the amount of molecular $^1H_2^+$ that overlaps with the deuterium peak.

A clear Sn enrichment of up to approximately 2 at.% is observed at the deuteride/α reaction front. The profile also indicates partitioning of the Sn between the two deuterides and preference for the Sn to partition to the α−matrix. We propose that these observations suggest the following qualitative statements:

- Sn solubility is higher in the α−matrix than in the deuteride phases
- Deuteride growth rate is controlled by the partition of Sn to the α-Zr matrix which leads to the formation of the Sn spike at the reaction front.

This fits well with the recent *ab inito* studies by Christensen et al. [35] that suggest of all the solute additions present in Zircaloy-4, Sn has the highest destabilizing effect on hydride phases and that Sn is relatively more stable in pure Zr than in any zirconium hydride. It follows that an enrichment of Sn at the reaction front as well as the need for Sn to leave the deuteride may retard further deuteride growth and have a significant influence on the growth rate. However, conversely, higher Sn content in Zr alloys increases surface oxidation and a recent trend among fuel manufacturers has been to reduce Sn content in Zr cladding material for reactors [36]. More research is needed on the effects Sn content and distribution has on the oxidation and hydride formation mechanisms combined and APT provides a valuable tool for investigating the segregation and partitioning behaviour of this critical solute addition.

Based on our body of evidence, we conclude that the grain boundary deuteride packets present in deuterium-charged Zircaloy-4 samples contain a network of δ and ζ with a rough ζ/α-Zr matrix growth front. The exact mechanism of formation of the ζ deuteride remains unclear. Possibly, δ grew during annealing at 400 °C and has partially transformed into the ζ phase during cooling while the amount of deuterium in the δ phase was increased at lower temperatures, as suggested by the phase diagram proposed by Grosse et al. [3]. Another possibility is that ζ formed first during the annealing as an intermediate transition phase and remained at the reaction front to accommodate the stresses between δ and α-Zr, in closer agreement with that proposed by Li et al. [14] and Shen et al. [16]. The FFT analysis of high resolution STEM images (Fig. 2 d) confirm the presence of the two deuteride phases. From the APT results, accurate quantification of D was challenging. Nevertheless, the 1D concentration profile (Fig. 4 b) shows a distinct change in relative D content consistent with the presence of multiple deuteride phases and the α-Zr matrix. The 1D concentration profile within the APT reconstruction at the growth front also showed a partitioning effect of Sn to the α-Zr matrix. The behaviour of Sn may have a significant influence over deuteride growth and is perhaps the rate-limiting factor due to the need for Sn to partition to the α-Zr matrix for further growth. Such information could have significant implications for controlling hydride evolution in Zr based alloys and reducing the severity of delayed hydride cracking.

Acknowledgements:

The first three authors equally contributed to the making of this article. SW prepared the alloy and performed the electrochemical charging as well as the optical microscopy and EBSD work. AB and BG led the writing of this letter. Uwe Tezins and Andreas Sturm are thanked for their support to the FIB and APT facilities at MPIE. AB, MH, DR, BG acknowledge the Deutsche Forschungsgemeinschaft (DFG) for partially funding this research through SFB 761 'Stahl ab initio'. AB also acknowledges the Alexander von Humboldt Foundation (AvH) for partially funding this research. The authors are grateful for the Max-Planck Society and the BMBF for the funding of the Laplace and the UGSLIT projects respectively. TBB thanks the Royal Academy of Engineering for support of his Research Fellowship. TBB and SW acknowledge support from the HexMat programme grant (EP/K034332/1). SW acknowledges Dr Vivian Tong for assistance with generating the EBSD unit cell orientations.

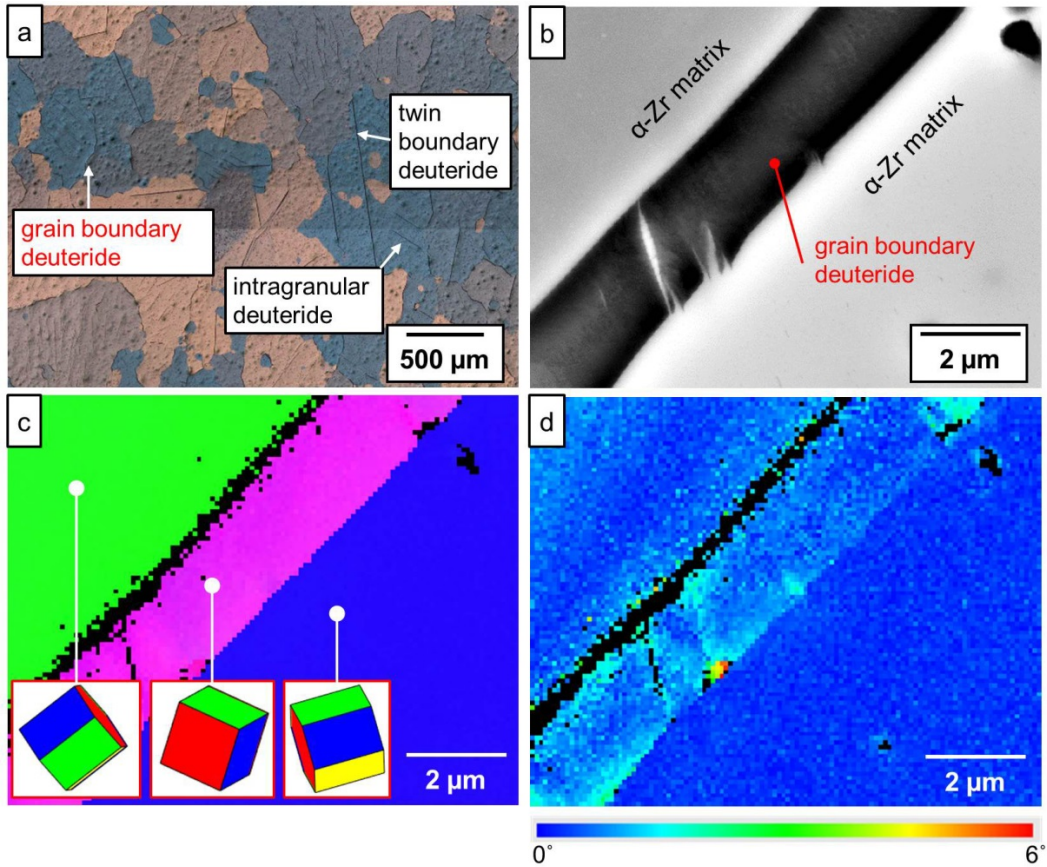

**Fig. 1** (a) Polarised light micrograph of Zircaloy-4 after deuterium charging showing evidence of deuteride formation (b) SEM-ECCI image of the same grain boundary deuteride indicated in (a). (c) relative crystal orientation EBSD map and unit cell orientations of the detected δ and α phases. (d) Misorientation to grain average orientation map. IPF: Inverse pole figure.

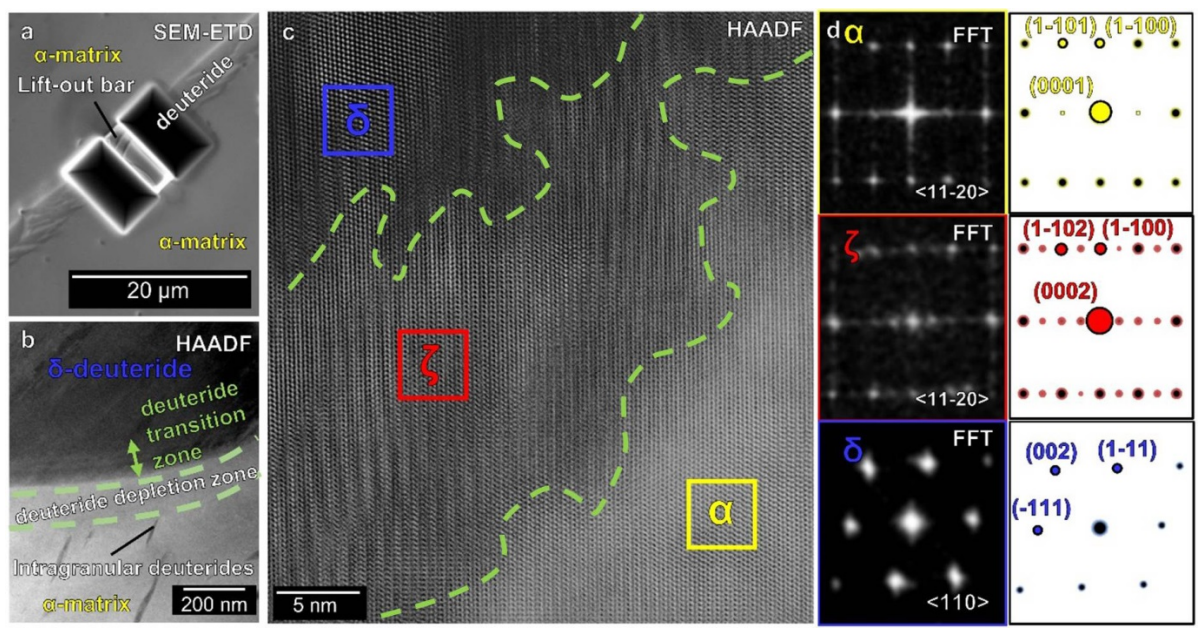

**Fig. 2** (a) SEM image of TEM lift-out location on grain boundary hydride of interest (b) STEM-HAADF micrograph showing a transitional region between matrix and hydride. (c) Detail HAADF micrograph of interfacial region showing hydride structure. (d) FFT of different regions in (c) with corresponding simulated diffraction patterns confirming different crystal structures of the α, ζ and δ regions.

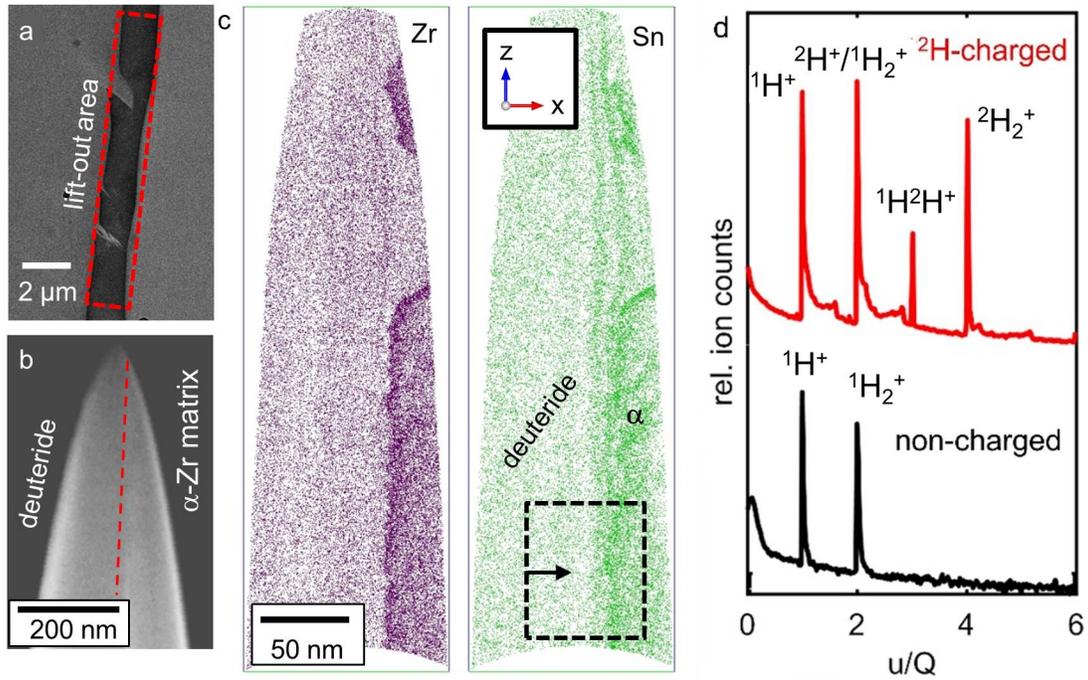

**Fig 3.** (a) APT lift-out region at grain boundary deuteride. (b) SEM image of specimen prior to APT experiment. The blue arrow indicates the suspected hydride/α interface. (c) APT reconstruction atom map cross-sections (5 nm thick). The region of interest and direction for the ID composition profile in Figure 4(b) is shown in the Sn map. (d) mass spectrum of dataset (red) which clearly shows the presence of peaks indicating D (or $^2$H) content after charging, a reference mass spectrum (black) from a non-charged specimen has also been included for comparison.

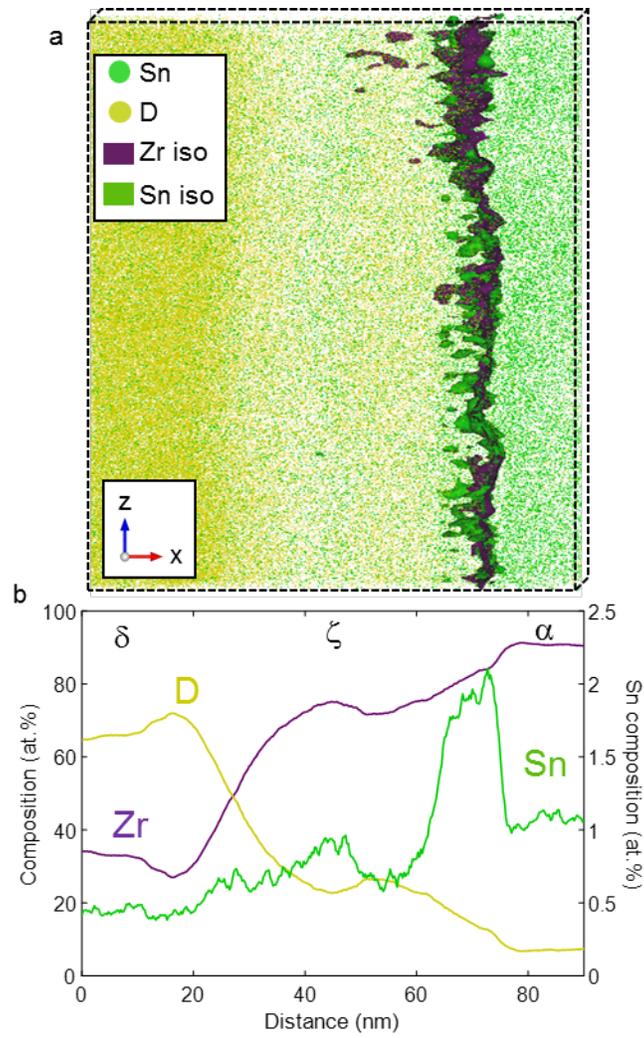

**Fig 4.** (a) zoomed in insert of region of interest across deuteride/α interface. A Zr isocomposition surface (85 at.%) and a Sn isocomposition surface (1.9 at.%) have been used to highlight the interface. (b) 1D composition profile of Sn in the x-direction.